\documentstyle[preprint,tighten,aps,amstex,graphicx,times,floats]{revtex}

\begin{document}

\thispagestyle{empty}

% macros for marking changes
\marginparwidth 1.cm
\setlength{\hoffset}{-1cm}
\newcommand{\mpar}[1]{{\marginpar{\hbadness10000%
                      \sloppy\hfuzz10pt\boldmath\bf\footnotesize#1}}%
                      \typeout{marginpar: #1}\ignorespaces}
\def\mda{\mpar{\hfil$\downarrow$\hfil}\ignorespaces}
\def\mua{\mpar{\hfil$\uparrow$\hfil}\ignorespaces}
\def\mla{\marginpar[\boldmath\hfil$\rightarrow$\hfil]%
                   {\boldmath\hfil$\leftarrow $\hfil}%
                    \typeout{marginpar: $\leftrightarrow$}\ignorespaces}

\def\ba{\begin{eqnarray}}
\def\ea{\end{eqnarray}}
\def\bq{\begin{equation}}
\def\eq{\end{equation}}

\renewcommand{\abstractname}{Abstract}
\renewcommand{\figurename}{Figure}
\renewcommand{\refname}{Bibliography}

% peter's conventions 
\newcommand{\eg}{{\it e.g.}\;}
\newcommand{\ie}{{\it i.e.}\;}
\newcommand{\etal}{{\it et al.}\;}
\newcommand{\ibid}{{\it ibid.}\;}

% additional commands 
\newcommand{\mx}{M_{\rm SUSY}}
\newcommand{\pt}{p_{\rm T}}
\newcommand{\et}{E_{\rm T}}
\newcommand{\del}{\varepsilon}
\newcommand{\sla}[1]{/\!\!\!#1}
\newcommand{\fb}{\;{\rm fb}}
\newcommand{\gev}{\;{\rm GeV}}
\newcommand{\tev}{\;{\rm TeV}}
\newcommand{\abi}{\;{\rm ab}^{-1}}
\newcommand{\fbi}{\;{\rm fb}^{-1}}

\preprint{
\font\fortssbx=cmssbx10 scaled \magstep2
\hbox to \hsize{
\hskip.5in \raise.1in\hbox{\fortssbx University of Wisconsin - Madison}
\hfill\vtop{\hbox{\bf MADPH-02-1275}
            \hbox{\today}                    } }
}

\title{ 
Charged Higgs Boson Production in Bottom-Gluon Fusion 
} 

\author{
Tilman Plehn 
} 

\address{ 
Physics Department, University of Wisconsin, Madison, WI 53706, USA \\
} 

\maketitle 

\begin{abstract}
  We compute the complete next-to-leading order SUSY-QCD corrections
  for the associated production of a charged Higgs boson with a top
  quark via bottom-gluon fusion. We investigate the
  applicability of the bottom parton description in detail. The higher
  order corrections can be split into real and virtual corrections for
  a general two Higgs doublet model and into additional massive
  supersymmetric loop contributions. We find that the perturbative
  behavior is well under control. The supersymmetric contributions
  consist of the universal bottom Yukawa coupling corrections and
  non-factorizable diagrams. Over most of the relevant supersymmetric
  parameter space the Yukawa coupling corrections are sizeable, while
  the remaining supersymmetric loop contributions are negligible.
\end{abstract} 

\vspace{0.2in}

%%%%%%%%%%%%%%%%%%%%%%%%%%%%%%%  MAIN TEXT  %%%%%%%%%%%%%%%%%%%%%%%%%%%%

\section{Higgs Physics at the LHC}

In the near future the CERN Large Hadron Collider (LHC) will be the
appropriate tool to look for physics beyond the Standard Model and to
determine its properties. The capabilities of the LHC beyond being a
pure discovery machine become increasingly important at energy scales
which are hard to access at a Linear Collider. 

The combined LEP precision measurements~\cite{lep_precision} suggest
the existence of a light Higgs boson. In the case of a single Standard
Model Higgs boson the LHC promises multiple coverage for any Higgs
boson mass, which will enable us to measure its different decay modes
and extract the couplings~\cite{lhc_smhiggs,lhc_tautau,lhc_tth}. In
the case of a supersymmetric Higgs sector this coverage becomes less
impressive. This is a direct consequence of the structure of the Higgs
sector: while the Minimal Supersymmetric Standard Model (MSSM)
predicts a light Higgs boson it also predicts an enhancement of the
coupling to down-type fermions, at the expense of the branching
fractions to gauge bosons.  This enhancement is an outcome from the
two Higgs doublet structure in the MSSM: one doublet is needed to give
mass to the up-type, the other one to the down-type fermions. The
vacuum expectation values of the two doublets are different,
parameterized by $\tan \beta = v_2/v_1$. The Yukawa coupling to the
down-type fermions is essentially enhanced by $\tan \beta$, while the
coupling to up-type fermions is suppressed by the same factor. In
addition to a light scalar Higgs boson the two Higgs doublet model
includes a heavy scalar, a pseudoscalar, and a charged Higgs boson.
None of these additional particles have a mass bounded from above,
apart from triviality or unitarity bounds. On the contrary, for a
large pseudoscalar mass these three additional particles all become
heavy and almost mass degenerate.\medskip

As in the Standard Model case the light scalar Higgs supersymmetric
boson will be produced via gluon fusion or weak boson fusion, but it
will most prominently decay to bottom quarks and tau leptons. A search
for the tau lepton decay essentially covers the MSSM parameter space
with a luminosity of $\sim 40 \fbi$ at the LHC~\cite{lhc_tautau}. The
same process can be used to determine if the light Higgs boson is the
scalar or the pseudoscalar mode in the two Higgs doublet model and
what kind of operator governs its coupling to gauge
bosons~\cite{lhc_vvh}. More exotic scenarios might for example lead to
an invisibly decaying light Higgs boson, which again can be extracted
from the backgrounds~\cite{lhc_invisible}.

All these observables linked to properties of a light Higgs boson can
serve as a probe if a new scalar particle is consistent with the
Standard Model Higgs boson. There is, however, only one way to
conclusively tell the supersymmetric Higgs sector from its Standard
Model counterpart: to discover the additional heavy Higgs bosons and
determine their properties. This task might entirely be left to
the LHC, since at a Linear Collider the promising production channels
are pair production of these heavy bosons, for which a first
generation collider might well have insufficient energy~\cite{tesla}.
At the LHC the possible enhancement of down-type fermion Yukawa
couplings renders the search for a heavy scalar and pseudoscalar Higgs
boson decaying to muon and tau lepton pairs most
promising~\cite{lhc_heavyhiggs}. For the charged Higgs boson the
coupling to fermions is more complex: for small values of $\tan \beta$
it is governed by the up-type coupling $m_u/\tan\beta$, whereas for
larger values of $\tan\beta$ the down type Yukawa coupling $m_d
\tan\beta$ dominates. In particular for values $\tan\beta \gtrsim 30$
the charged Higgs coupling behaves the same way as the heavy neutral
Yukawa couplings.  While the chances of finding a heavy Higgs boson
with a small value of $\tan\beta$ at the LHC are rather slim, the
discovery of all heavy Higgs scalars in the large $\tan\beta$ regime
is likely.\bigskip

Three search modes for the charged Higgs boson have been explored in
some detail: (1) Charged Higgs bosons can be pair produced in a
Drell--Yan type process, mediated by a weak interaction
vertex~\cite{pairs_dy}. Moreover, they can be pair produced at tree
level in bottom quark scattering~\cite{pairs_bb} or through a one loop
amplitude in gluon fusion~\cite{pair_gg}. (2) A charged Higgs boson
can be produced together with a $W$ boson via scattering of two bottom
quarks or in gluon fusion~\cite{higgs_w_asso}. (3) The charged Higgs
boson can be produced in association with a top quark, which seems to
be the most promising search
channel~\cite{charged_theo,charged_nlo,charged_zhu}.  The charged
Higgs boson can be detected either decaying to a top and a bottom
quark~\cite{charged_top} or decaying to a tau lepton and a
neutrino~\cite{charged_tau}. The completely exclusive process reads
\footnote{There is an additional contribution from $q\bar{q}$
  scattering, where the charged Higgs boson is produced through
  intermediate $b\bar{b}$ or $t\bar{t}$ states. Numerically this
  contribution is negligible at the LHC. It is also irrelevant for the
  following discussion, where we are interested in incoming gluons
  splitting into two bottom quarks. Therefore we omit this process in
  our discussion of exclusive $\bar{b}tH^-$ production at the LHC.}:
\begin{equation}
gg \to \bar{b} t H^- + c.c. \qquad \qquad
H^- \to \tau \bar{\nu}_\tau \quad \text{or} \quad
H^- \to b \bar{t}
\label{eq:sig_excl}
\end{equation}

As we will argue in Section~\ref{sec:bottom_parton} this process can
and should be evaluated in the bottom parton approximation $bg \to
tH^-$, unless the observation of the additional bottom jet is
necessary to extract the signal out of the background. Recently both
LHC experiments have published detailed studies of this production
channel with very promising results~\cite{charged_atlas,charged_cms}.
However, the crucial ingredient to searches and in particular to the
precise extraction of couplings and masses at the LHC are
next-to-leading order predictions for the signal and background cross
sections.  Without these improved cross section calculations
theoretical uncertainties will almost immediately become the limiting
factor in many analyses. The next-to-leading order cross section
predictions for the inclusive production process $bg \to tH^-$ will be
presented in Section~\ref{sec:nlo} for a general two Higgs doublet
model and in Section~\ref{sec:susy} for the MSSM.\bigskip

\underline{Conventions:} Throughout this entire paper we show
consistent leading order or next-to-leading order cross section
predictions, including the respective one loop or two loop strong
coupling constant, running heavy quark masses, and the corresponding
CTEQ5L or CTEQ5M1 parton densities~\cite{cteq5}. The bottom pole mass
is fixed as $4.6\gev$, to give the correct $\overline{\rm MS}$ mass
$m_b(m_b)=4.2\gev$~\cite{running_mass}. We usually assume three
charged Higgs masses of $250, 500, 1000\gev$, and if not stated
otherwise $\tan\beta=30$. The exclusive cross sections are quoted with
a massive (4.6\gev) bottom quark in the matrix element and the phase
space, the inclusive results are evaluated for a vanishing bottom
mass. The bottom Yukawa coupling is set to the running bottom mass,
unless explicitly stated as being the pole mass.  When we talk about
the running bottom Yukawa coupling we implicitly include the running
top Yukawa coupling to the charged Higgs boson as well
($y_{b,t}(\mu_R)$), but the running of the bottom mass is the dominant
effect, by far. As the central value all scales are set to the average
final state mass $\mu=m_{\rm av}=(m_t+m_H)/2$. The extension of this
calculation to charged Higgs boson masses below the top mass is
straightforward: to avoid double counting of diagrams which also
appear in top pair production with a subsequent decay into a charged
Higgs boson and a bottom jet we will have to subtract on-shell top
states. This is the standard procedure for supersymmetric production
processes and can be applied to light charged Higgs boson production
without any modification~\cite{lhc_susy_strong,lhc_susy_weak}.

\section{Bottom Parton Scattering} 
\label{sec:bottom_parton}

As a starting point in this discussion we emphasize that the exclusive
production channel $gg \to \bar{b}tH^-$ is a perturbatively well
defined way to compute the total cross section as well as
distributions for associated $tH^-$ production. It is consistent in
the sense that it includes the squared matrix element to order
$\alpha_s^2 y_{b,t}^2$, where $y_{b,t}$ is the charged Higgs Yukawa
coupling to the third generation quarks. Even though there might be
some dispute concerning the precise numerical value of the bottom
quark mass, the infrared divergences arising from the
intermediate bottom quark propagators are regularized by this finite
bottom quark mass.  Once these bottom quarks are observed or even
tagged, the bottom quark transverse momentum and rapidity become the
relevant cutoff parameters to define the observable cross section
including the detector acceptance cuts; they render the cross
section after cuts almost independent of the actual value of the
bottom mass, which would be the relevant cutoff parameter for the
total cross section without acceptance cuts. \smallskip

Beyond naive perturbation theory the integration over phase space of
the final state bottom quark gives rise to possibly large
logarithms~\cite{bottom_parton}. As an illustration the typical gluon
radiation off an incoming parton in Drell--Yan production processes
leads to an asymptotic $1/p_{T,g}$ behavior in the gluon transverse
momentum distribution. The same problem arises in exclusive charged
Higgs boson production, where one of the two incoming gluons splits
into two bottom quarks, eq.(\ref{eq:sig_excl}). Because the massive
bottom propagator leads to an asymptotic transverse mass dependence
$1/m_{T,b}$ instead of the transverse momentum $1/p_{T,b}$, the
infrared divergence is regularized by the bottom mass. For small
transverse bottom momenta the differential partonic cross section
approaches the asymptotic form~\cite{spira_susy}
\begin{alignat}{7}
\frac{d \sigma^{(btH)}}{d p_{T,b}} &\sim
\frac{d \sigma^{(btH)}}{d p_{T,b}} \Bigg|_{\rm asympt} = 
S \; \frac{p_{T,b}}{m_{T,b}^2} =
S \; \frac{p_{T,b}}{p_{T,b}^2+m_b^2}  \notag \\
\sigma_{\rm tot}^{(btH)} &\sim
\sigma_{\rm tot}^{(btH)} \Big|_{\rm asympt} = 
\frac{S}{2} \; \log \left( \frac{\mu_F^2}{m_b^2} + 1 \right)
\label{eq:excl_asymp}
\end{alignat}
with a proportionality constant $S$, which we can link to the 
asymptotic total cross section. In contrast,
for large transverse momentum $p_{T,b} \gg m_b$ we can safely neglect
all bottom mass effects. The integration over the bottom phase space
leads to logarithms $\log(p_{T,b}^{\rm max}/m_b)$. They are not
divergent, but they can become quite large, though not as dramatically
as for light quarks where $\Lambda_{\rm QCD}$ serves as the infrared
cutoff.  Switching to a bottom quark parton description ($bg \to
tH^-$) corresponds to a resummation of these potentially large
logarithms beyond naive perturbation theory. However, this procedure
relies on several approximations, which should be carefully
examined.\bigskip

\begin{figure}[t] 
\begin{center}
\includegraphics[width=8.0cm]{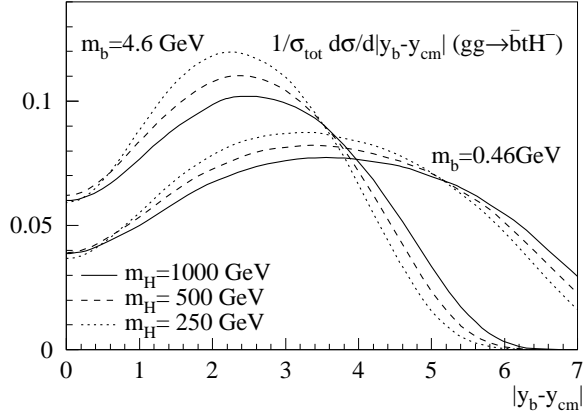}
\end{center}
\vspace*{0mm}
\caption[]{\label{fig:del_y_excl} 
  The rapidity difference between the final state bottom jet and the
  center of mass system for exclusive charged Higgs boson production
  at the LHC, eq.(\ref{eq:sig_excl}). The two sets of curves with
  three different charged Higgs boson masses are given for the
  physical on-shell bottom mass $4.6\gev$ as well as for an
  arbitrarily chosen smaller bottom mass as the infrared
  regulator.}
\end{figure}

When describing the intermediate bottom as a parton we use the DGLAP
evolution with the splitting kernels for massless particles.  We
assume that the bottom quark be massless.\footnote{This
  approximation does not have to include the bottom Yukawa coupling.
  We can consistently expand the cross section in terms of the bottom
  mass, extracting an over-all factor $y_{b,t}^2$ first. In other
  words, once we consider the Standard Model as an effective theory
  with massive fermions there is no link between the masses and the
  Yukawa couplings. This becomes obvious in the two Higgs doublet
  model, where we consistently neglect terms proportional to $m_b$,
  but keep terms proportional to $m_b (\tan\beta)^j \; (j\ge1)$.}  In
turn we also assume that at leading order the intermediate bottom
quark and therefore the outgoing bottom jet are collinear with the
incoming partons in the exclusive process. This approximation will
never be perfect, since the cutoff parameter $m_b$ is only slightly
smaller than the minimum observable transverse momentum at a collider.
But for the parton description of the bottom quark is it a necessary
condition that the outgoing bottom in the exclusive cross section is
clearly peaked forward. We show this behavior for exclusive charged
Higgs boson production in Fig.~\ref{fig:del_y_excl}. For the physical
bottom mass the distribution is indeed peaked forward, and as expected
the peak moves further out for smaller bottom masses. A detailed
discussion of the error induced by the zero bottom mass approximation
can be found in ref.~\cite{bottom_parton,neutral_bh}.\smallskip

\begin{figure}[t] 
\begin{center}
\includegraphics[width=8.0cm]{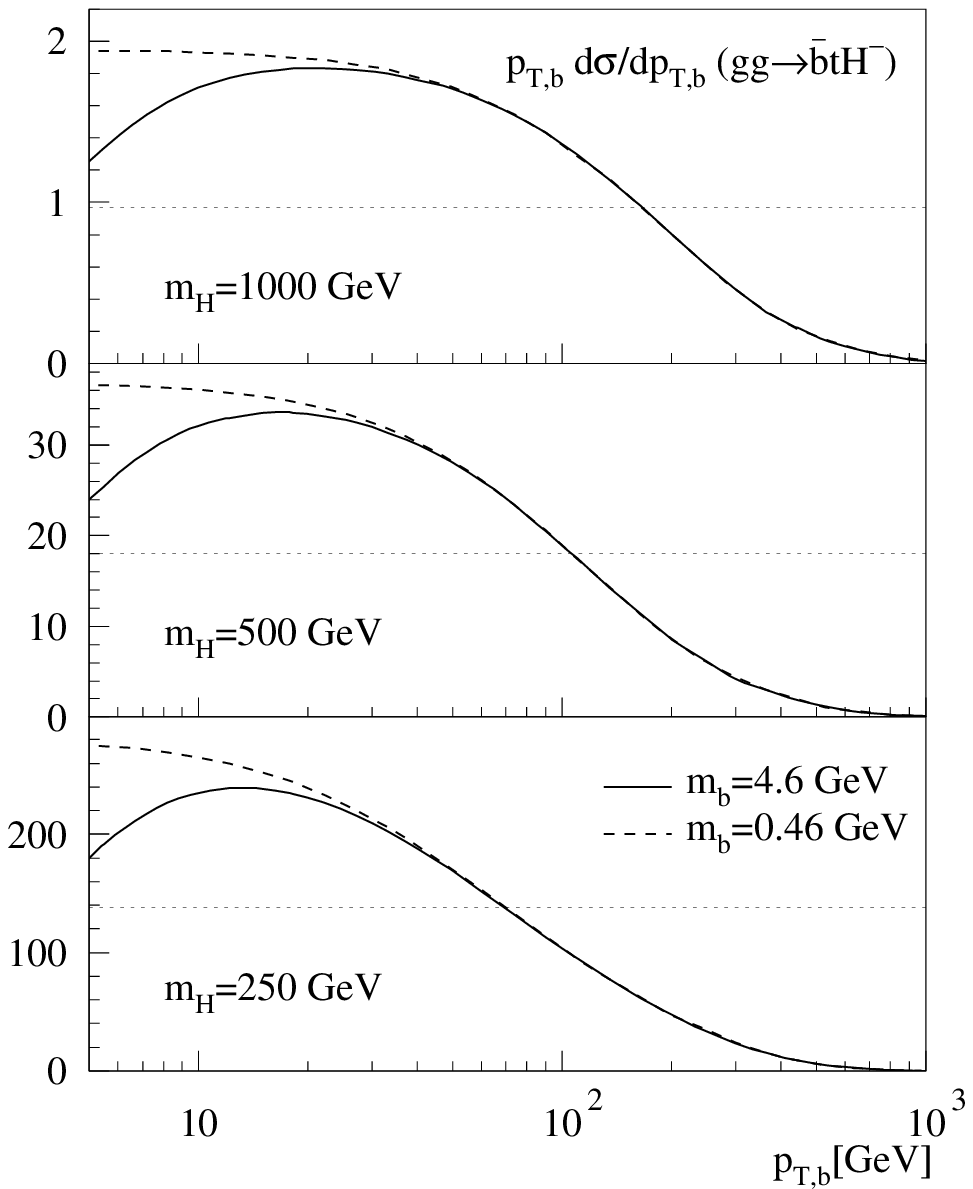} \hspace{10mm}
\includegraphics[width=8.0cm]{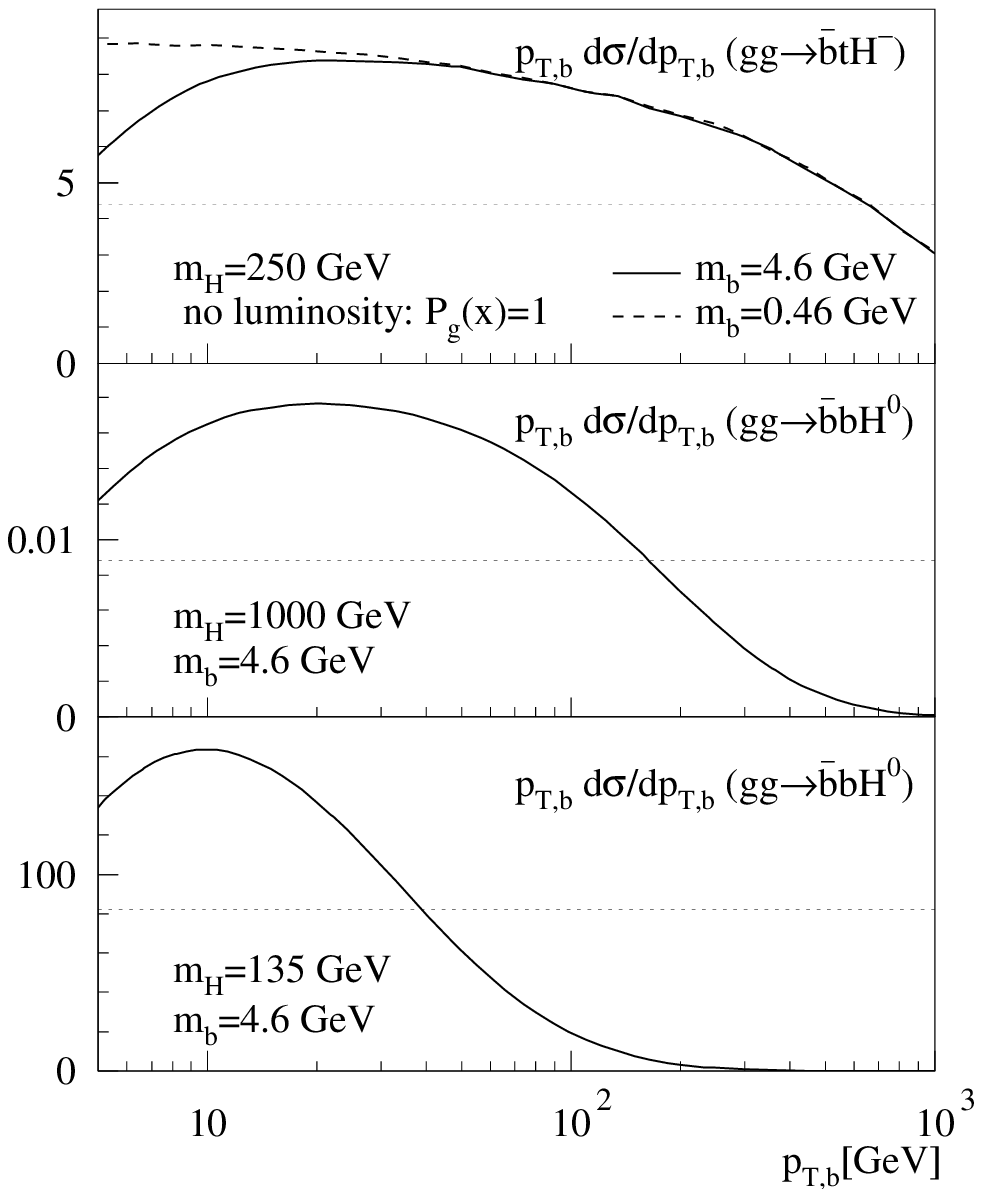}
\end{center}
\vspace*{0mm}
\caption[]{\label{fig:plateau} 
  Left: the bottom transverse momentum distribution for exclusive
  charged Higgs boson production at the LHC, eq.(\ref{eq:sig_excl}).
  For all three Higgs masses the curves are given for the physical
  on-shell bottom mass $4.6\gev$ as well as for an arbitrarily chosen
  smaller bottom mass as the infrared regulator. The thin dotted line
  indicates half the height of the plateau. The absolute normalization
  of the curves for the two infrared regulators is physical. Both
  curves coincide for large transverse momenta, where the bottom mass
  is negligible. Right: in the upper panel the same distribution for a
  heavy charged Higgs boson, but with the gluon luminosity set to
  unity ${\cal L}_{gg} \equiv 1$. Below this in the two lower panels
  the transverse momentum distribution for the bottom quarks in
  exclusive neutral Higgs boson production $gg \to \bar{b}bH$ for two
  neutral Higgs boson masses.}
\end{figure}

After making sure that the collinear approximation describes the
kinematics of the tree-level process $gg \to \bar{b}tH^-$ we still
have to determine if there are large logarithms to resum. In the
bottom parton approach we approximate the complete differential cross
section by an asymptotic $1/p_{T,b}$ or $1/m_{T,b}$ behavior.  The
upper boundary of the $p_{T,b}$ or $m_{T,b}$ integration defines the
factorization scale $\mu_F$ of the bottom parton density and
determines how big the resummed logarithms can be. After integrating
out the final state bottom quark in the exclusive gluon fusion
process the total hadronic cross section for $pp \to gb \to tH^-$
production becomes essentially proportional to $\log(\mu_F/m_b)$, as
we would expect. While this $1/m_{T,b}$ behavior is by definition
present even for large values of the transverse momentum in the matrix
element for the corresponding Feynman diagram, this is not necessarily
true for the differential hadronic cross section $d \sigma/d p_{T,b}$.

In Fig.~\ref{fig:plateau} we show the $1/p_{T,b}$ behavior of the
hadronic distributions for three different charged Higgs boson masses.
All renormalization and factorization scales are set to the average
final state particle mass. First of all we see how the zero bottom
mass approximation breaks down when the transverse momentum is of the
order of the bottom mass and a distinction between transverse mass and
transverse momentum is necessary. Instead of a simple $1/p_{T,b}$ we
indeed see the asymptotic form from eq.(\ref{eq:excl_asymp}). If we
replace the on-shell bottom mass with a smaller bottom mass the
plateau extends to smaller transverse momentum, again confirming the
asymptotic behavior. The small $p_{T,b}$ end of plateau in the
transverse momentum spectrum, however, does not lead to large
numerical effects, since the logarithm $\log(p_{T,b}^{\rm max}/m_b)$
and thereby the bottom parton density vanish for a factorization scale
$\mu_F=p_{T,b}^{\rm max} \sim m_b$.

Looking for large numerical effects we have to focus on the high
$p_{T,b}$ end of the asymptotic regime. In the left panel of
Fig.~\ref{fig:plateau} we see how the high $p_{T,b}$ end of the
plateau roughly scales with the average mass in the final state. This
coincides neatly with the observation that the only scales allowed for
the evaluation of total cross sections are external scales. They are
typically chosen proportional to the average mass of the final state
particles
\begin{equation}
 \mu_F \sim C \; m_{\rm av} = C \; \frac{m_t + m_H}{2}
\label{eq:m_av}
\end{equation}
where the proportionality factor $C$ is arbitrary. The dependence on
the choice of the scale and thereby on the choice of $C$ vanishes
after including all orders of perturbation theory.  Comparing
eq.(\ref{eq:m_av}) with Fig.~\ref{fig:plateau} shows that the naive
choice $C \sim 1$ is not obviously appropriate. Choosing $C \sim 1$
assumes large logarithms $\log(p_{T,b}^{\rm max}/m_b)$ being resummed
to values $\mu_F \sim m_{\rm av}$. This will yield an overestimate of
the total cross section.\medskip

Using the asymptotic form of the cross section in
eq.(\ref{eq:excl_asymp}) we first note that the value of $S$ should
only very mildly depend on the numerical value of the bottom
mass~\cite{bottom_parton,spira_susy}. The same is true for the
factorization scale, which only parameterizes the large transverse
momentum regime. We can see from Fig.~\ref{fig:plateau} that there the
bottom mass effects are negligible. Evaluating the expression for the
asymptotic total cross section for the two bottom masses we can
determine the values of $S$ and $\mu_F$. As a check we compare the
value of $S$, which is the predicted plateau value of $p_{T,b}
d\sigma/dp_{T,b}$, with the plateau value we obtain from the complete
calculation. We find them good agreement at least for a bottom mass of
$0.46\gev$ in the case where the plateau is not particularly well
pronounced for the physics bottom mass. For the appropriate
factorization scale we obtain $185, 120, 80\gev$ for the three Higgs
boson masses $1000, 500, 250\gev$. Very similar values we would
naively obtain from Fig.~\ref{fig:plateau}, looking for the point
where $p_{T,b} d\sigma/dp_{T,b}$ has dropped to half of the plateau
value. This means that the appropriate factorization scale indeed
scales with the average final state mass, eq.(\ref{eq:m_av}), but with
$C \sim 1/3$. On the other hand we point out that for associated
charged Higgs boson and top quark production we do always find a
$p_{T,b}$ regime in which the hadronic differential cross section $d
\sigma/d p_{T,b}$ shows the expected asymptotic behavior and therefore
the bottom parton treatment is justified --- but with an appropriate
choice of the bottom parton factorization scale.\medskip

To understand where this unexpectedly narrow asymptotic plateau comes
from we turn to the partonic cross section. In the right panel of
Fig.~\ref{fig:plateau} we show the transverse momentum distribution
with the gluon luminosity set to unity (${\cal L}_{gg} \equiv 1$).
Still the interference between the different diagrams as well as the
hadronic phase space limit the asymptotic behavior once we look at
very large transverse momenta. If one would want to determine the
bottom factorization scale for example from the $p_{T,b}$ value at
which the plateau has dropped to half of its value, we find $\mu_F
\sim 3 m_{\rm av}$ when we discard the gluon luminosity. The preferred
low scales observed from the left panel of Fig.~\ref{fig:plateau} are
therefore entirely due to the steeply falling gluon density which
suppresses any large transverse momentum radiation of forward bottom
jets.\medskip

To prove the universality of our argument we show the same transverse
bottom momentum distribution for the exclusive neutral Higgs boson
production $gg \to b \bar{b} H$~\cite{neutral_lhc} in the right panel
of Fig.~\ref{fig:plateau}. This channel becomes important for large
values of $\tan \beta$, where it supplements the inclusive Higgs
production process via gluon fusion~\cite{lhc_smhiggs,lhc_heavyhiggs}.
It can of course be evaluated as an exclusive process with incoming
gluons $gg \to \bar{b}bH$. But it can also be regarded as
partly~\cite{neutral_bh} or completely inclusive, \ie with one or two
incoming bottom partons. The numerical effects of the resummation in
the bottom parton approach can be as large as an order of magnitude
for the total cross section.  The same reasoning as for the charged
Higgs boson production applies in this case.  First one shows that the
bottom quarks are forward or for a small bottom mass collinear to the
incoming gluons.  Then one determines an appropriate choice of the
factorization scale from the size of the asymptotic region in which
the differential cross section shows the behavior as in
eq.(\ref{eq:excl_asymp}). One has to keep in mind that the expected
asymptotic behavior (once it does not give a plateau in
Fig.~\ref{fig:plateau}) shows non-negligible bottom mass effects.
Therefore we emphasize that for differential cross sections at leading
order the bottom parton approximation is not valid if the regime
where the finite bottom mass ruins the $1/p_{T,b}$ behavior
immediately blends into the regime where the gluon densities cut off
the asymptotic behavior at large transverse momentum.

From the comparison of the two curves for a $1\tev$ neutral and a
$1\tev$ charged Higgs boson we see that the behavior is very similar:
the bottom parton description is valid, and the factorization scale
should be chosen considerably below the average final state mass (for
the charged Higgs boson) or below the Higgs mass (for the neutral
Higgs boson). For a light neutral Higgs boson ($m_H=135\gev$) the
asymptotic behavior only survives up to $p_{T,b} \lesssim 40\gev$, in
more detail depending on where one would like to draw the line. This
corresponds to a logarithmic enhancement $\log(p_{T,b}/m_b) \lesssim
\log 8 \sim 2$. Even more so for the Tevatron this leads to
factorization scales where the bottom parton density decreases, and
with it the enhancement of the total cross section, which is the
effect of the resummation.\footnote{As we will show later, higher
order QCD contributions to the inclusive processes~\cite{neutral_bh}
include the exclusive channel $gg \to b \bar{b} H$. For very small
factorization scales this exclusive diagram becomes dominant and leads
us back to the original exclusive cross section in a well defined
manner, once we consider the next-to-leading order cross
section.}\bigskip

\begin{figure}[t] 
\begin{center}
\includegraphics[width=8.0cm]{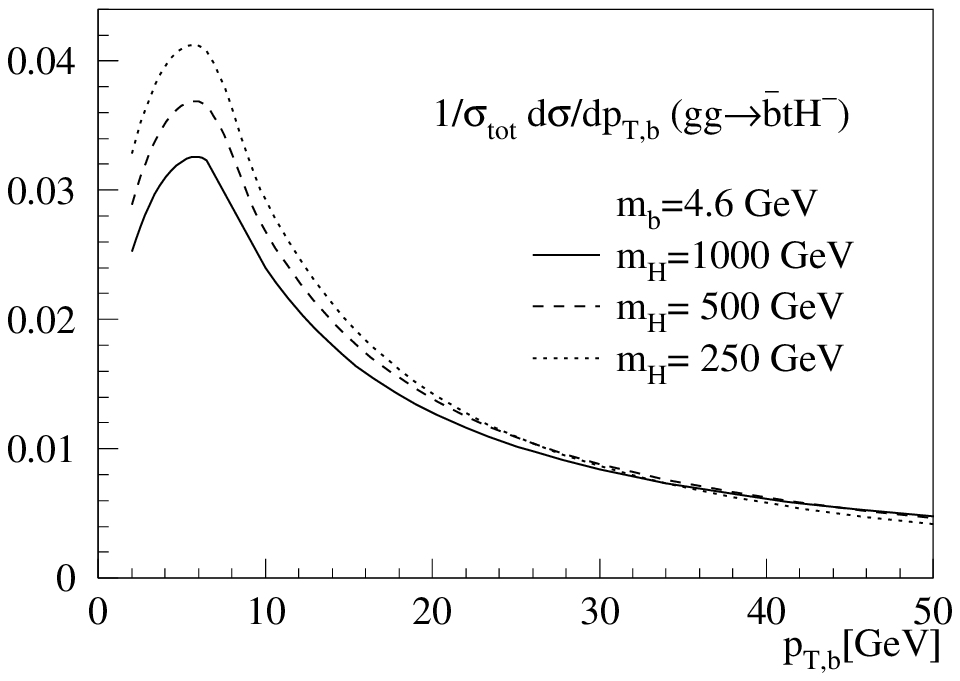} \hspace{10mm}
\includegraphics[width=8.0cm]{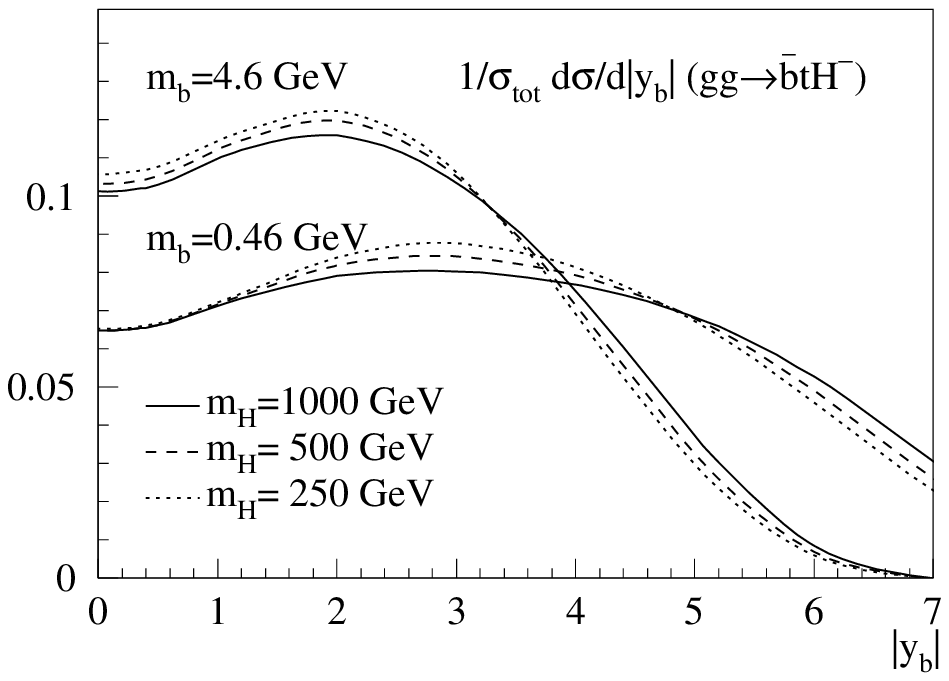}
\end{center}
\vspace*{0mm}
\caption[]{\label{fig:excl_ypt} 
  Left: the bottom transverse momentum distribution for exclusive
  charged Higgs production at the LHC, eq.(\ref{eq:sig_excl}).
  Right: the bottom rapidity distribution for the same process. Again
  a set of curves with a small infrared regulator is added
  ($m_b=0.46\gev$).}
\end{figure}

Up to this point we have only talked about the validity of the bottom
parton approximation and the correct choice of the factorization
scale. However, the applicability of the bottom parton approach is
very closely tied to the reason why the partly inclusive analyses are
attractive: if the exclusive process exhibits a collinear final state
bottom jet from gluon splitting this jet is not likely to hit the
detector, much less to be tagged. For the exclusive $\bar{b}tH^-$
production we illustrate this feature in Fig.~\ref{fig:excl_ypt}. Most
of the bottom jets are not sufficiently central to be tagged and
thereby significantly suppress the backgrounds. Moreover, the bottom
transverse momentum peaks around $p_{T,b} \sim m_b$, considerably too
soft to be seen or even tagged with good efficiency. This means that
the same feature which allows us to use the bottom parton approach
makes it hard to utilize the exclusive process: the final state bottom
jet is too collinear to be particularly useful.

Even though the exclusive cross section with the appropriate cuts ---
but without a required final state bottom jet --- yields a well
defined perturbative cross section prediction, the presence of
collinear bottom jets can lead to large logarithms. They alter the
convergence of the strictly perturbative power series. Therefore the
inclusive process with the right choice of parameters gives a
numerically improved cross section prediction. In the case in which
the analysis does not require a final state bottom jet we strongly
advocate use of the inclusive process, since the reliability of the
cross section predictions will be significantly improved beyond naive
perturbation theory.

\section{Next-to-leading Order Results for a Two Higgs Doublet Model}
\label{sec:nlo}

To improve the theoretical cross section prediction and to reduce the
theoretical uncertainty we compute the inclusive process $pp \to gb
\to tH^-$ to next-to-leading order QCD. In this Section we present the
results for a two Higgs doublet model. We would like to mention that
part of the numbers presented in this section have been compared in
detail with similar results given earlier in
ref.~\cite{charged_zhu}. For all diagrams included in both
calculations the numbers agree within the uncertainties from different
input parameters and from the scheme dependence in the top mass
renormalization. The complete set of next-to-leading order QCD
corrections include virtual gluon loops as well as real gluon
radiation. The massive supersymmetric loops will be discussed in
Section~\ref{sec:susy}. The complete set of next-to-leading order
processes consists of:
\begin{alignat}{7}
g b \;       &\to \; t H^- \qquad \qquad \text{(Born term)}          \notag \\
g b \;       &\to \; t H^- \qquad \qquad \text{(virtual correction)} \notag \\
g b \;       &\to \; t H^- g                                  \notag \\
g g \;       &\to \; t H^- \bar{b}                            \notag \\
q \bar{q} \; &\to \; t H^- \bar{b} \qquad \qquad
b \bar{q} \;  \to \; t H^- \bar{q} \qquad \qquad
b \bar{b} \;  \to \; t H^- \bar{b}                            \notag \\
b      q  \; &\to \; t H^- q \qquad \qquad
b      b \;   \to \; t H^- b  
\label{eq:nlo_proc}
\end{alignat}

The calculation is carried out in the dimensional regularization
scheme. All ultraviolet poles are analytically cancelled between the
virtual diagrams and the counter terms. The strong coupling and the
bottom Yukawa coupling are renormalized in the $\overline{\rm MS}$
scheme.  This way $\alpha_s$ and $y_{b,t}$ both are running
parameters, dependent on the same renormalization scale $\mu_R$. As
the renormalization scale we choose $\mu_R = m_{\rm av}$. We expect
logarithms from virtual corrections to be absorbed in the running mass
definition, in complete analogy to Higgs decays to massive
fermions~\cite{higgs_decay_nlo}.  The factorization and the
renormalization scales are often identified for convenience, but there
is no argument from first principles which enforces that choice. We
will discuss this issue in detail below. The external top mass we
renormalize in the on-shell scheme.\smallskip

\begin{figure}[t] 
\begin{center}
\includegraphics[width=8.0cm]{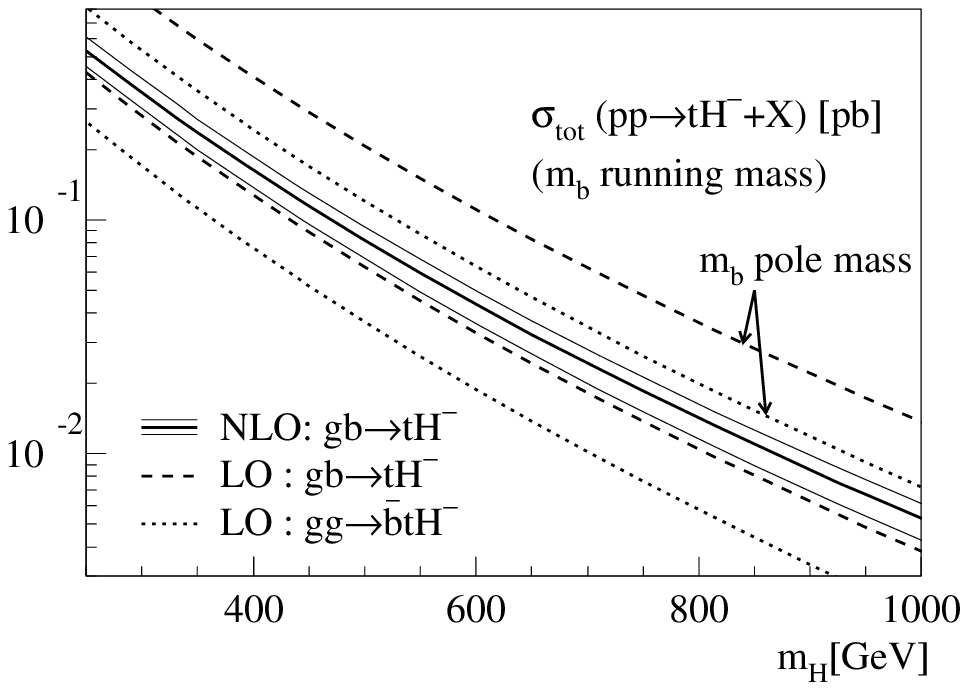} \hspace{10mm}
\includegraphics[width=8.0cm]{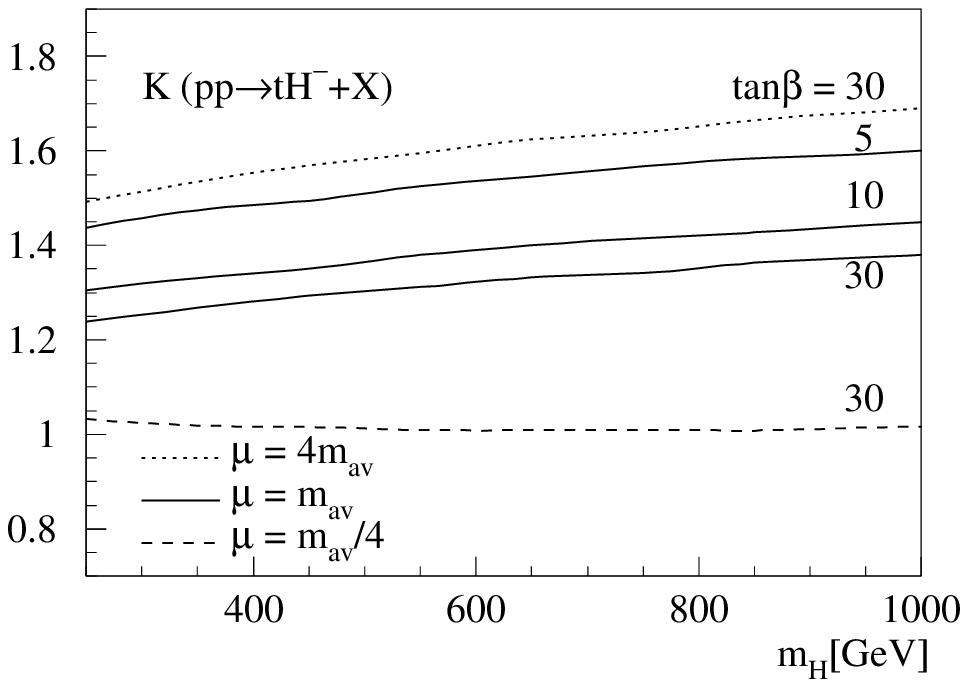}
\end{center}
\vspace*{0mm}
\caption[]{\label{fig:incl_sig} Left: the inclusive production cross
  section $pp \to tH^-+X$ at the LHC. The dashed and solid lines show
  the consistent leading order and next-to-leading order results. The
  dotted line is the total cross section from the exclusive production
  process, eq.(\ref{eq:sig_excl}). To illustrate the enhancement
  through large logarithms both tree level results are also quoted
  using the (inappropriate) pole mass for the bottom Yukawa
  coupling. The range for the next-to-leading order result is given
  for $\mu_F=\mu_R=m_{\rm av}/4 \cdots 4 m_{\rm av}$. Right: the
  corresponding consistent $K$ factors for the three values of
  $\tan\beta=5,10,30$. In the case of $\tan\beta=30$ we show three
  choices of $\mu = \mu_R = \mu_F$, consistently for leading order and
  next-to-leading order cross sections.}
\end{figure}

The infrared poles are also cancelled analytically between the virtual
corrections, the real emission diagrams, and the mass factorization.
The numerical impact of the higher order contributions is shown in
Fig.~\ref{fig:incl_sig}. The leading order results are given for the
running bottom mass as well as for the bottom pole mass in the Yukawa
coupling. This choice is not fixed by first principles at leading
order, whereas at next-to-leading order the counter term defines the
bottom Yukawa coupling uniquely. The difference between these two mass
definitions is strictly speaking part of the theoretical uncertainty
for the leading order cross section prediction. After adding all
higher orders the cross section should be independent of the choice,
as it should be independent of the renormalization and factorization
scale. We want to stress, however, that it is well known that the pole
mass Yukawa coupling always yields a huge overestimate of cross
sections and decay widths and should generally not be
used~\cite{higgs_decay_nlo}. The band for the next-to-leading
order cross section is given by a variation of the renormalization and
factorization scale $\mu_R=\mu_F=(m_{\rm av}/4, m_{\rm av}, 4 m_{\rm
av})$. From the discussion in Section~\ref{sec:bottom_parton} we know
that for the factorization scale this is not a good choice. But we
still fix the two scales for convenience at the central scale, which
is preferred by the renormalization
scale~\cite{higgs_decay_nlo}.\medskip

The size of the next-to-leading order corrections as a function of the
charged Higgs boson mass and of the scale is shown in the right panel
of Fig.~\ref{fig:incl_sig}. The $K$ factor is defined consistently as
$\sigma_{\rm NLO}/\sigma_{\rm LO}$, including the respective one or
two loop running of the strong coupling and the third generation
Yukawa couplings. The corrections seem to be perturbatively well under
control, ranging from $+30\%$ to $+40\%$ for $\tan\beta=30$ and Higgs
boson masses between $250$ and $1000\gev$. As expected the size of the
$K$ factor still depends on the choice of the scales.

In addition to the explicit $K$ factor, the shift in the consistent
bottom Yukawa coupling absorbs another factor $y^2_{b,{\rm
2-loop}}/y^2_{b,{\rm 1-loop}} \sim 0.84$, while the top Yukawa
coupling is more stable\footnote{ We could in principle use the 3-loop
running bottom masses, which yields another factor $y^2_{b,{\rm
3-loop}}/y^2_{b,{\rm 2-loop}} \sim 0.97$.  The physical condition is
again $m_b(m_b)=4.2\gev$.  However, this way we would resum and absorb
terms which are not explicitly included in the NLO cross section and
the actual numerical improvement is not obvious and certainly not well
under control.}, $y^2_{t,{\rm 2-loop}}/y^2_{t,{\rm 1-loop}} \sim
1.0$. The next-to-leading order QCD corrections are flavor blind and
proportional only to the Born coupling structure $y_{b,t}^2$, which as
a function of $\tan\beta$ is either dominated by the top quark or by
the bottom quark Yukawa coupling. In the right panel of
Fig.~\ref{fig:incl_sig} we show the $K$ factor for three different
values of $\tan\beta$ (the curve for $\tan\beta=50$ is
indistinguishable from $\tan\beta=30$).  The only difference between
these three curves comes from the running Yukawa coupling: the running
bottom Yukawa coupling, which is dominant for large values of
$\tan\beta$, absorbs a larger correction than the running top Yukawa
coupling. The consequence is a larger remaining $K$ factor for smaller
values of $\tan\beta$.\bigskip

\begin{figure}[t] 
\begin{center}
\includegraphics[width=8.0cm]{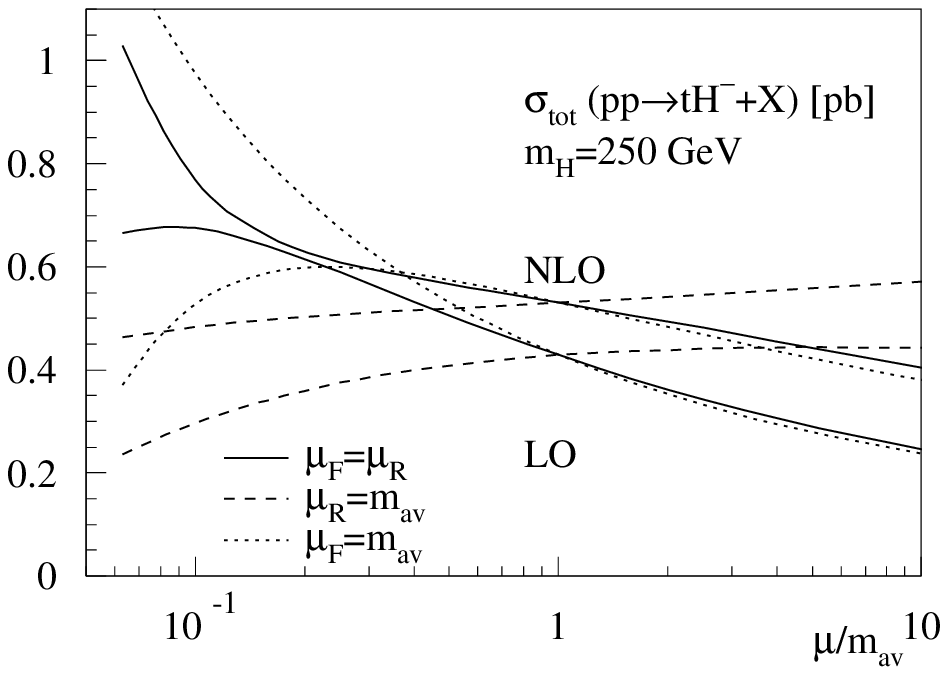} \hspace{10mm}
\includegraphics[width=8.0cm]{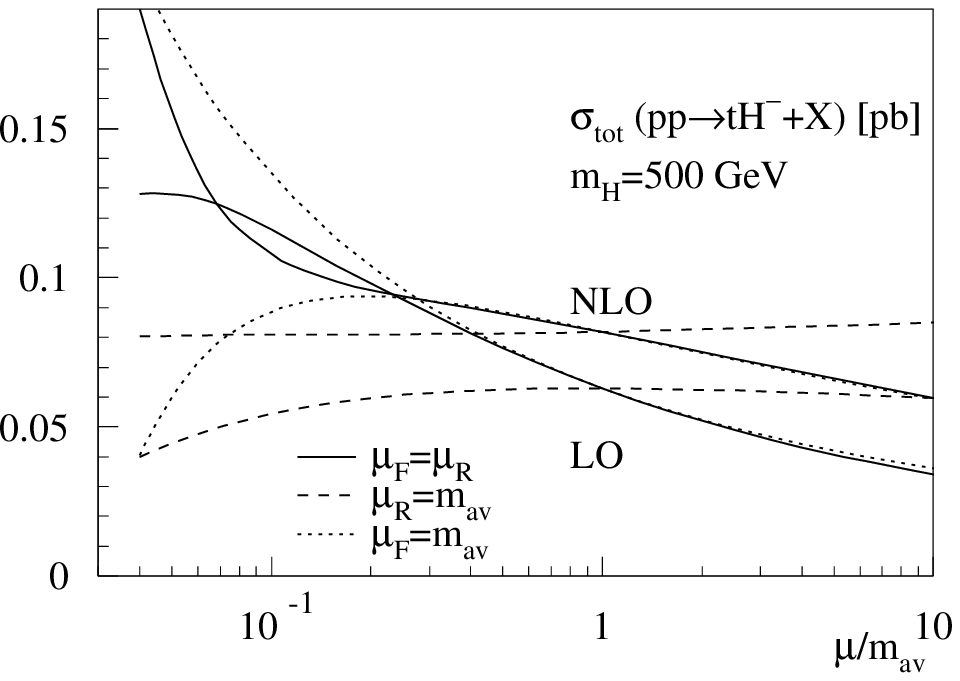}
\end{center}
\vspace*{0mm}
\caption[]{\label{fig:incl_scale} 
  The variation of the total inclusive cross section $pp \to tH^-+X$
  as a function of the renormalization and factorization scales,
  around the central value $\mu=m_{\rm av}$, eq.(\ref{eq:m_av}).  The
  two panels give the result for two different charged Higgs boson
  masses, $250\gev$ and $500\gev$. The lower end of the curves
  corresponds to $\mu \sim 10\gev$. The respective leading order and
  next-to-leading order curves can be identified at the point where
  they meet for the central choice $\mu = m_{\rm av}$.}
\end{figure}

More detailed information concerning the scale variation is included
in Fig.~\ref{fig:incl_scale}. As argued above, the appropriate choice
for the factorization scale should scale with the average final state
mass, but with a proportionality factor smaller than unity $\mu_F \sim
1/3 \; m_{\rm av}$. In the discussion of the total cross section
results we accommodate this effect by choosing a large window for the
scale variation. From Fig.~\ref{fig:incl_scale} we see that the
dependence of the cross section on the factorization scale is mild. To
leading order the dependence becomes large only once the bottom
factorization scale comes close to the bottom mass. Since the bottom
density comes from gluon splitting into two bottom quarks it has to be
essentially proportional to $\log(\mu_F/m_b)$, \ie it has to vanish
for $\mu_F \to m_b$. This is precisely the behavior we see in the
small scale regime for both Higgs boson masses. To next-to-leading
order the scale dependence stays flat even for very small
factorization scales. Assuming that the light flavor quark initiated
processes listed in eq.(\ref{eq:nlo_proc}) are suppressed at the LHC
the purely gluon initiated process dominates for factorization scales
$\mu_F \to m_b$.  The large $K$ factor is an artifact of the bottom
parton approximation which leads to a vanishing leading order cross
section, whereas the next-to-leading order saturates onto the
light-flavor induced channels, which include the exclusive $gg
\to \bar{b}tH^-$ process.  This way the next-to-leading order
inclusive calculation interpolates between the inclusive and the
exclusive results\footnote{It remains to be checked, however, how good
  this interpolation is numerically in the regime where the `large
  logarithms' $\log(p_{T,b}/m_b)$ are only slightly enhanced.},
where now the exclusive channel does not depend on the bottom mass as
the infrared regulator. Instead all infrared poles cancel in the given
order of perturbation theory. This means that at the one-loop level
the inclusive cross section approaches the exclusive result in the
limit of no large logarithms, where the enhancement through the
resummation disappears. The only error left is the zero bottom mass
approximation~\cite{neutral_bb,neutral_bh}.\bigskip

Once the charged Higgs boson is heavier than $\sim 500\gev$ the
numerically dominant theoretical uncertainty comes from the unknown
renormalization scale, dominantly from the scale of the strong
coupling. While for a small factorization scale the total cross
section decreases, a small renormalization scale yields a larger
strong coupling and a larger running bottom mass. Identifying both
scales inherently leads to a cancellation and therefore to a likely
underestimate of the theoretical uncertainty. This can for example be
taken care of by identifying a large renormalization scale with a
small factorization scale~\cite{higgs_nnlo}.\medskip

On the other hand, the physics can easily be understood. For small
factorization scales the cross section decreases slowly, until the
factorization scale becomes close to the bottom mass, at which point
it drops sharply. This reflects the logarithmic dependence of the
bottom parton density. At next-to-leading order the drop is softened
by the light-flavor induced channels, in particular with a purely
gluonic initial state. At large scales the logarithmic dependence
$\sim \log(\mu_F/m_b)$ is still present, but the variation has become
very weak.\smallskip

The renormalization scale dependence in contrast explodes for the
leading order cross section at small scales long before reaching the
bottom mass.  At next-to-leading order it reaches a maximum, but the
variation of the cross section is still considerably larger than the
variation with the factorization scale. The cancellation between the
renormalization and the factorization scale dependence has an
interesting consequence, which we observe in
Fig.~\ref{fig:incl_scale}. If we identify both scales and evaluate the
cross section for very small values $\mu/m_{\rm av}\lesssim 0.1$ the
next-to-leading order prediction increases rapidly. Physically this is
not a problem, since the scales have to be very small, which might be
an appropriate choice for the factorization scale, but certainly not
for the renormalization scale, as we argued above. We know that for
these small scales the dependence on the logarithms $\log(\mu_F/m_b)$
and $\log(\mu_R/m_H)$ largely cancels. However, terms proportional to
$\log(\mu_F/m_b) \times \log(\mu_R/m_H)$ in particular in the $gg$
channel can become very large. One way to look at this effect is that
the unphysically small renormalization scale gives a large negative
prefactor for the factorization scale dependence, namely
$\log(\mu_R^2/m_H^2)$. This dominates the factor in front of
$\log(\mu_F/m_b)$, which for more appropriate renormalization scales
is small and positive instead.\medskip

For a reasonably large renormalization scale almost the entire scale
variation is driven by the renormalization scale; \ie over almost the
entire range the renormalization scale dominates the variation of the
cross section with the scales. This effect is well known from
supersymmetric particle production at the LHC.  For processes mediated
by a strong coupling at tree level, the scale variation is an
appropriate measure for the theoretical
uncertainty~\cite{lhc_susy_strong}. Again the change in the cross
section is driven by the renormalization scale. On the other hand, for
weakly interacting particles produced in Drell--Yan type processes,
the leading order scale variation is dominated by the factorization
scale and is not a good measure for the theoretical
uncertainty~\cite{lhc_susy_weak}. For the inclusive associated charged
Higgs boson and top quark production both, Fig.~\ref{fig:incl_sig} and
Fig.~\ref{fig:incl_scale} show that the remaining theoretical
uncertainty as derived from the renormalization and factorization
scale dependence can be estimated to be $\lesssim 20\%$ for a central
choice of scales.\medskip

To accommodate this behavior we stick to the identification of both
scales $\mu_F=\mu_R= C \; m_{\rm av}$, as defined in
eq.(\ref{eq:m_av}). According to Fig.~\ref{fig:incl_scale} this
reflects the dominant scale variation of the cross section. In
addition we follow our arguments of Section~\ref{sec:bottom_parton}
and check that the cross section predictions are stable for small
factorization scales, down to at least $\mu_F \sim m_{\rm av}/3$. The
graphs in Fig.~\ref{fig:incl_scale} confirm that the cross sections
are stable down to factorization scales $\mu_F \lesssim m_{\rm
  av}/10$, which also means that the inclusive charged Higgs boson
production for $m_H \gtrsim m_t$ will not run into any problems with
the bottom parton description.\bigskip

\begin{figure}[t] 
\begin{center}
\includegraphics[width=8.0cm]{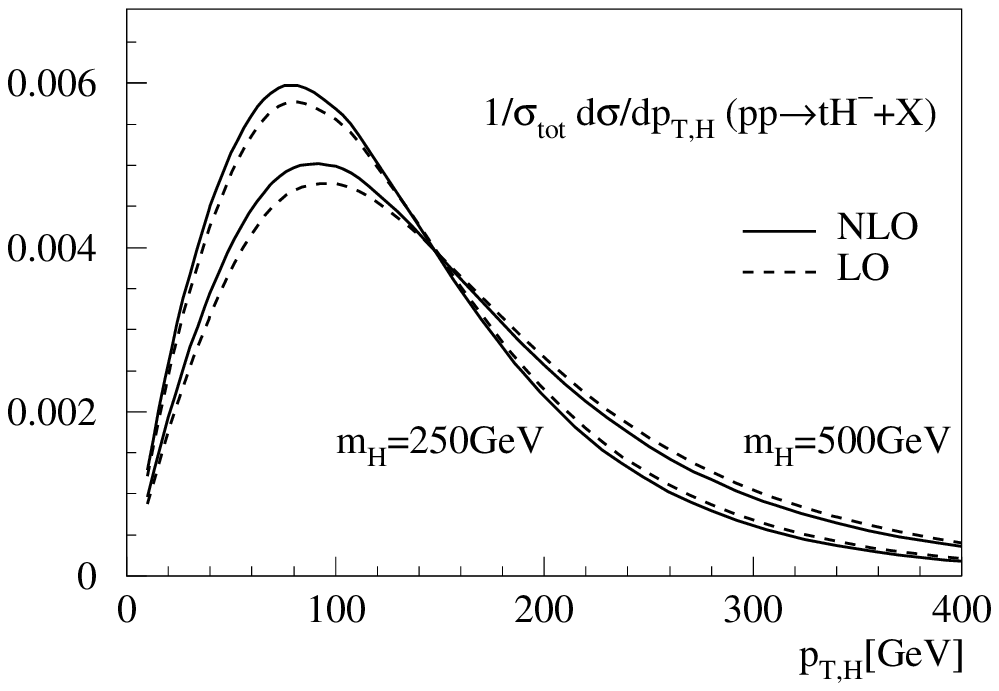} \hspace{10mm}
\includegraphics[width=8.0cm]{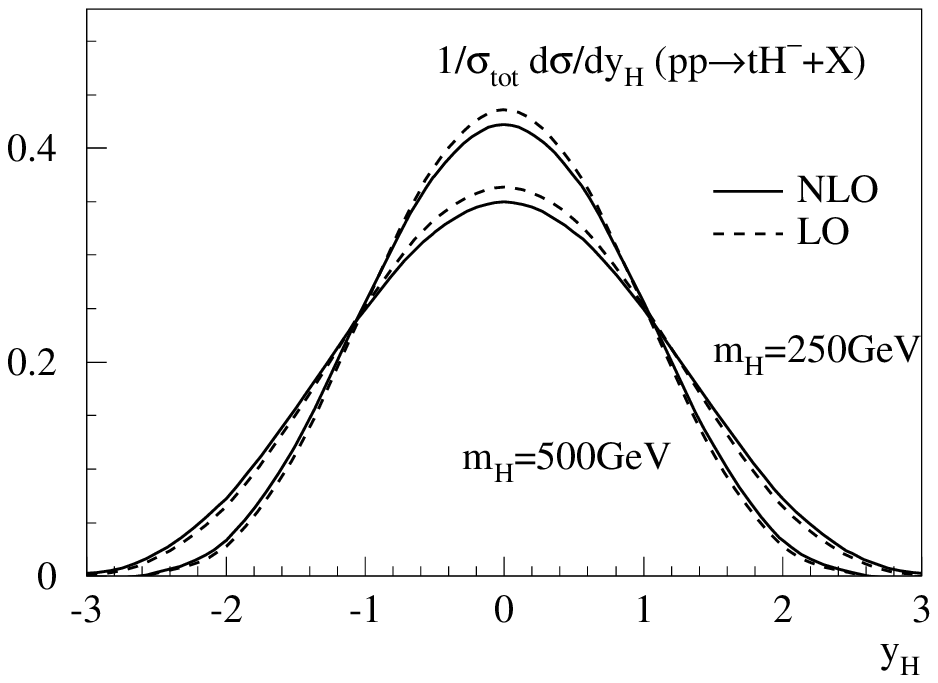}
\end{center}
\vspace*{0mm}
\caption[]{\label{fig:incl_ypt} 
  The charged Higgs boson transverse momentum and rapidity
  distributions for the inclusive process $pp \to tH^-+X$ are given
  for two different charged Higgs boson masses, $250\gev$ and
  $500\gev$. The distributions are normalized to the total cross
  section and evaluated at the central scale $\mu = m_{\rm av}$.}
\end{figure}

Looking beyond the corrections to the total hadronic cross section we
compute the transverse momentum and rapidity distributions for the
charged Higgs boson. The normalized differential cross sections are
depicted in Fig.~\ref{fig:incl_ypt}. As expected, the impact of the
higher order corrections on the shape of the rapidity distribution is
small; the addition of the third final state particle does not alter
the symmetric behavior around $y_H=0$ at a $pp$ collider. The effect
on the charged Higgs boson transverse momentum is a systematic
softening. One might have expected slightly harder charged Higgs
bosons, with an additional gluon radiated off the top quark and both
of them balanced by the Higgs boson. However, most of the jet
radiation comes from the initial state. As seen in
Section~\ref{sec:bottom_parton}, the radiation of high transverse
momentum jets is cut off by the steeply falling partonic energy
dependence of the gluon luminosities.  This limited available energy
directly translates into a softening of the Higgs boson transverse
momentum, once a third final state particle is added to the process.

\section{Next-to-leading Order Results with Supersymmetry}
\label{sec:susy}

Even though the Standard Model with a two doublet Higgs sector is a
perfectly well-defined renormalizable theory, we are particularly
interested in the MSSM version of this model. The MSSM fixes the
parameters of the Higgs sector, links each of the Higgs doublets to
up- or down-type fermions, normalizes the two gauge couplings to the
Fermi coupling constant, and fixes all three- and four-scalar
couplings.  The number of free tree level parameters in the Higgs
sector is reduced to two, which are usually chosen to be the
pseudoscalar mass $m_A$ and the ratio of the vacuum expectation values
$\tan\beta$~\cite{hhg}.\bigskip

At next-to-leading order, supersymmetric particles can propagate
through loops and contribute to the cross section $bg \to tH^-+X$.
Because supersymmetry is broken and all virtual particles are heavy
these corrections are infrared finite. The ultraviolet poles have to
be extracted and absorbed into supersymmetric contributions to the
counter terms for bare Standard Model masses and coupling. All
next-to-leading order corrections to the total cross section coming
from these supersymmetric loop diagrams we include in a supersymmetric
correction factor
\begin{equation}
  K_{\rm SUSY} = \frac{\sigma_{\rm SUSY} + \sigma_{\rm NLO}}{\sigma_{\rm NLO}}
               = 1 + \frac{1}{K} \; \frac{\sigma_{\rm SUSY}}{\sigma_{\rm LO}}
\label{eq:ksusy}
\end{equation} \medskip

As in Section~\ref{sec:nlo} we assume a massless bottom quark.  In
supersymmetry this adds a slight complication: the bottom squark mass matrix
includes off-diagonal elements, which are parameterized as $-m_b
(A_b+\mu \tan\beta)$. The splitting of the first term $m_b A_b$ into
the bottom mass and a trilinear mass parameter is not enforced by the
Lagrangean; in other words the combination $m_b A_b$ does not
automatically have to vanish with a zero bottom quark mass. Similarly
in the approximation of zero bottom mass $m_b$ on the one hand and finite
bottom Yukawa coupling $m_b \tan\beta$ on the other, this 
off-diagonal matrix element will not vanish either. The
off-diagonal term induces a mixing between the supersymmetric partner
of the left and right handed bottom quark: we have to work with mass
eigenstates $\tilde{b}_{1,2}$ instead of interaction eigenstates
$\tilde{b}_{L,R}$ even in the limit of a vanishing bottom mass.\medskip

\begin{figure}[t] 
\begin{center}
\includegraphics[width=8.0cm]{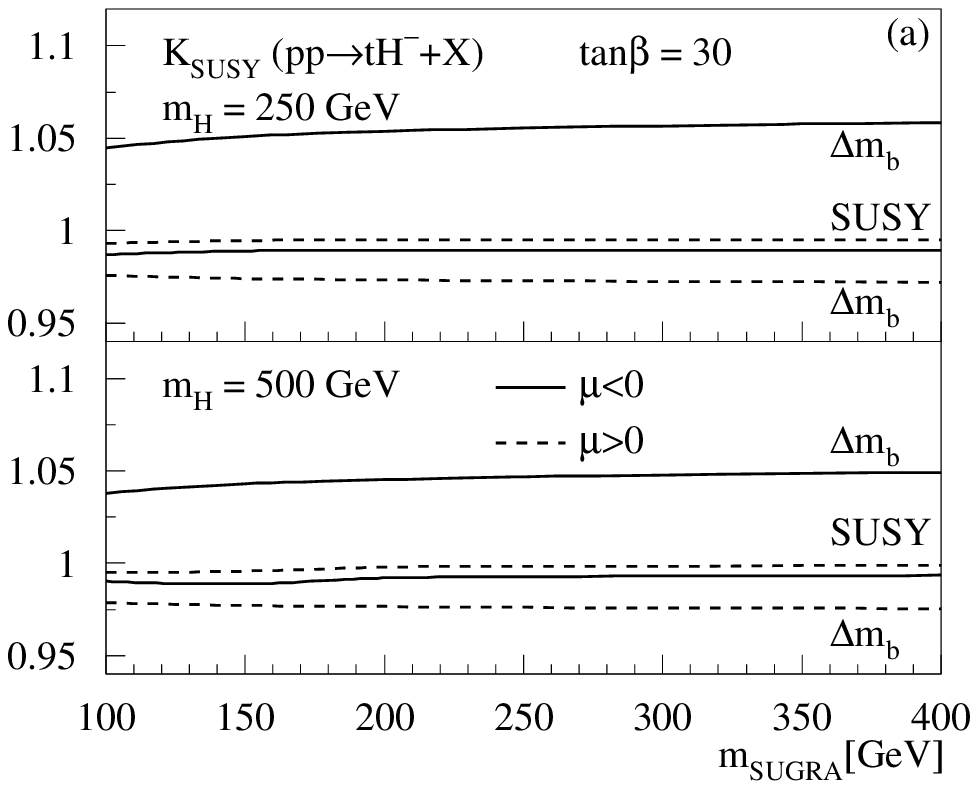} \hspace{10mm}
\includegraphics[width=8.0cm]{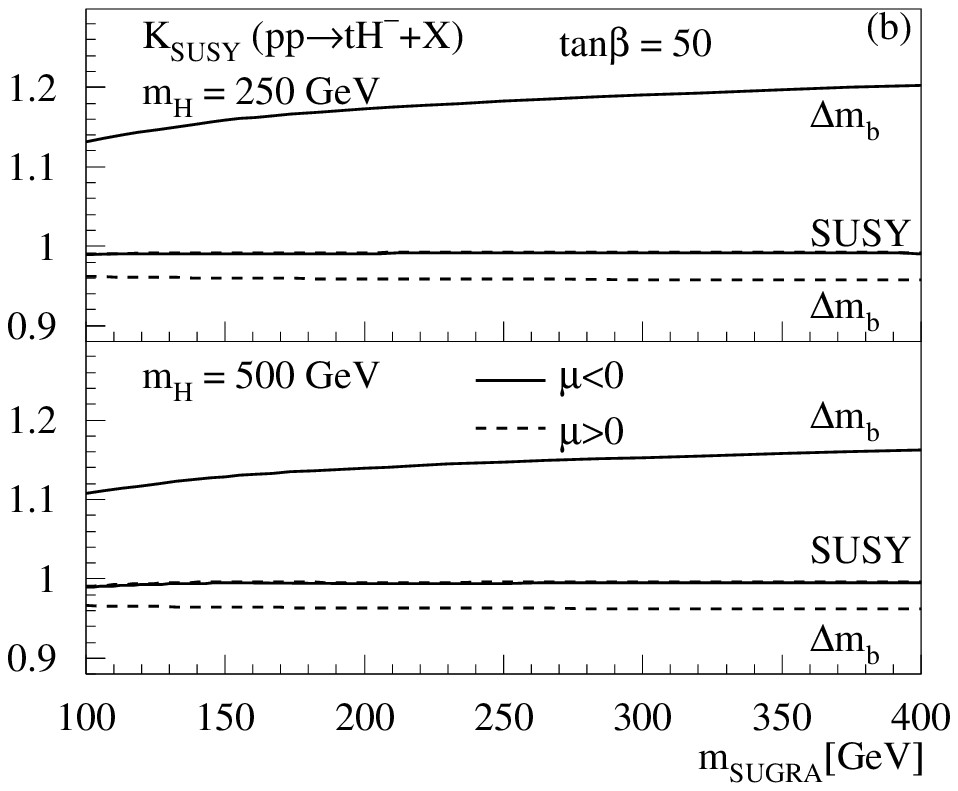} \\[4mm]
\includegraphics[width=8.0cm]{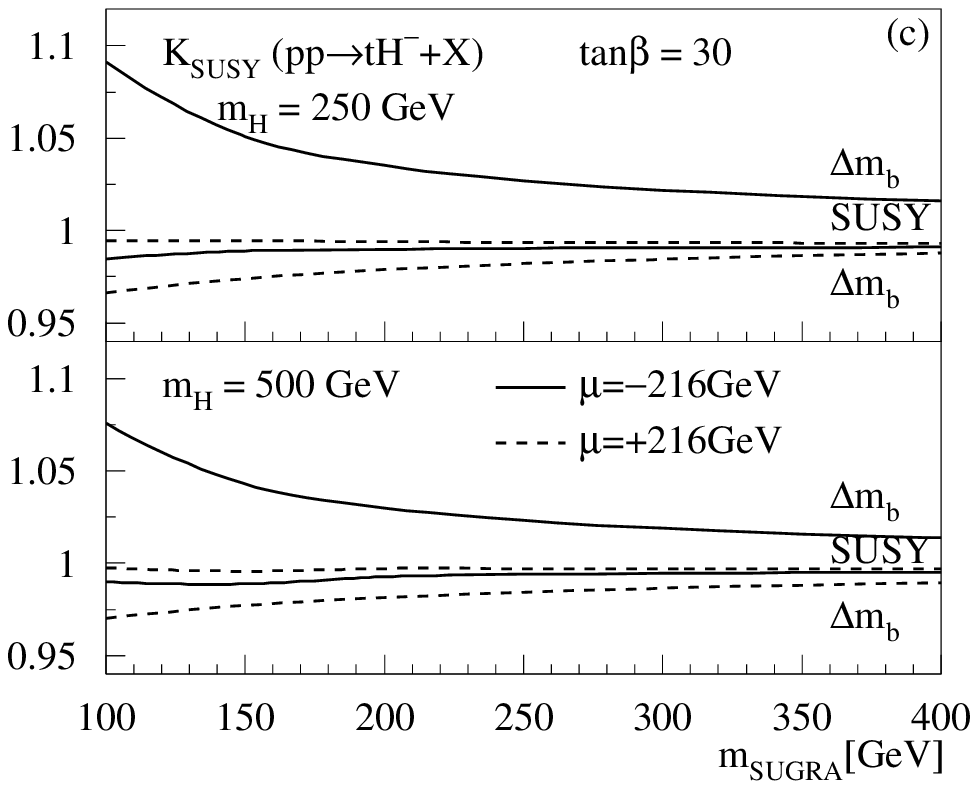} \hspace{10mm}
\includegraphics[width=8.0cm]{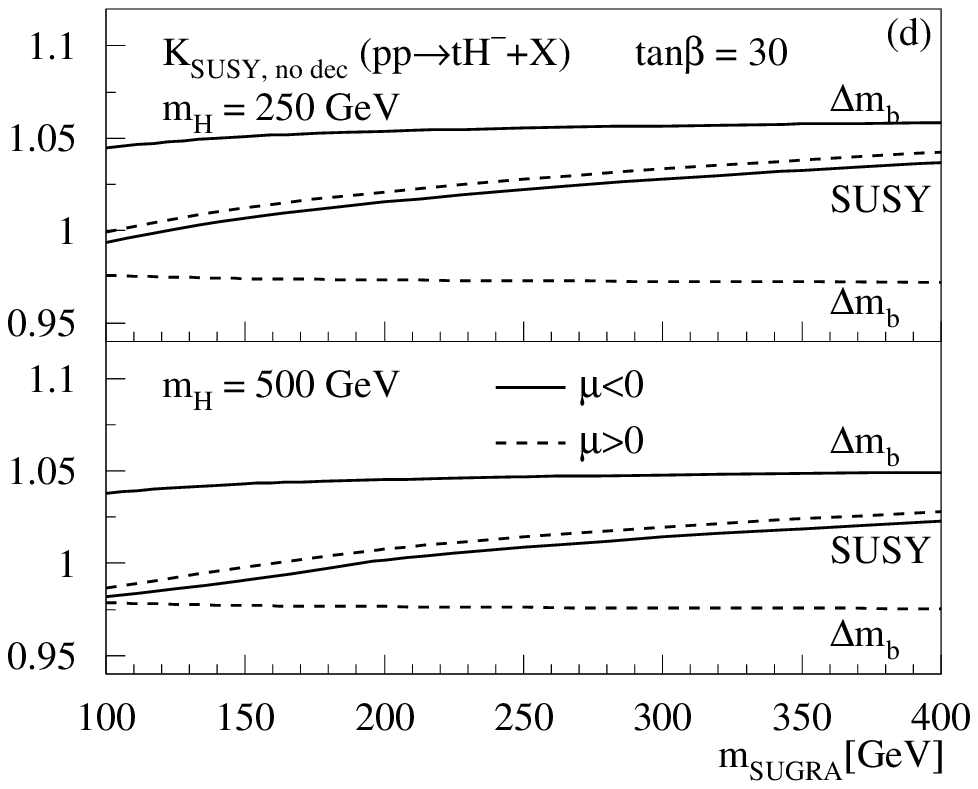}
\end{center}
\vspace*{0mm}
\caption[]{\label{fig:susy}
  The dependence of the total cross section $pp \to tH^-+X$ on
  supersymmetric loop contributions. The enhancement factor is defined
  in eq.(\ref{eq:ksusy}). The curves with the typically larger
  deviation from the two Higgs double model include $\Delta m_b$
  corrections only, the curves labeled SUSY include the complete
  remaining set of loop supersymmetric diagrams. All curves are given
  for two different Higgs boson masses (upper/lower panel) and for two
  signs of $\mu$ (solid/dashed line). The mass scale is defined as
  $m_{\rm SUGRA}=m_0\equiv m_{1/2}$: (a) corrections for
  $\tan\beta=30$ and with a running higgsino mass parameter $\mu$; (b)
  same as (a), but with $\tan\beta=50$; (c) same as (a), but with
  $\mu$ fixed at its value for $m_{\rm SUGRA}=150\gev$; (d) same as
  (a), but without decoupling the heavy spectrum from the running
  Yukawa coupling. The range of particle masses covered by $m_{\rm
    SUGRA}=100...400\gev$ are for the gluino mass $284...1017\gev$,
  for the sbottom masses $212...827\gev$ and $265...901\gev$, and for
  the stop masses $199...687\gev$ and $326...895\gev$. The higgsino
  mass parameter runs from $|\mu|=136...595\gev$, except for in part
  (c).}
\end{figure}

At one-loop order this off-diagonal entry can connect a left handed
with a right handed bottom quark. Even though in the final result we
neglect the bottom mass we do have to take into account
this contribution to the bottom mass counter term.  Mass counter terms
have to be proportional to the bare mass $\delta m_b \propto m_b$; in
this special case we find that in the on-shell mass renormalization
scheme $\delta m_b \propto \sin(2 \theta_b)$, with an implicit
dependence $\sin(2 \theta_b) \propto m_b (A_b+\mu \tan\beta)$. This
gives back the proportionality to the bare mass, but as argued
above it means that the contribution to the mass counter term has to
be kept even in the zero bottom mass approximation. As shown in a
series of papers this mass counter terms modifies the relation between
the bottom mass and the bottom Yukawa
coupling~\cite{delta_mb,delta_mb_resum}:
\begin{alignat}{7}
\frac{m_b \tan\beta}{v}  \; &\to \; \frac{m_b \tan\beta}{v} 
                           \; \frac{1}{1+\Delta m_b} \notag \\ 
\Delta m_b &= \;    \frac{\sin(2 \theta_b)}{m_b} 
                 \; \frac{\alpha_s}{4 \pi} \; C_F \; m_{\tilde{g}}
                 \; \frac{1}{i \pi^2}
                 \; \left[ B(0,m_{\tilde{b},2},m_{\tilde{g}})
                          -B(0,m_{\tilde{b},1},m_{\tilde{g}})
                    \right] \phantom{haaallllooooooo} \notag \\
           &= \; \frac{\alpha_s}{4 \pi} \; C_F \; m_{\tilde{g}} 
              \; \left(A_b + \mu \tan\beta \right) 
              \;  I(m_{\tilde{b},1},m_{\tilde{b},2},m_{\tilde{g}}) 
                                \notag \\[3mm]
I(a,b,c)   &= - \frac{1}{(a^2-b^2)(b^2-c^2)(c^2-a^2)} \;
                \left[ a^2b^2 \log \frac{a^2}{b^2}
                      +b^2c^2 \log \frac{b^2}{c^2}
                      +c^2a^2 \log \frac{c^2}{a^2} \right]
\label{eq:delta_mb}
\end{alignat}
The functions $B(0,m_{\tilde{b}},m_{\tilde{g}})$ are the usual scalar
two-point functions with the integration measure $d^nq$, $C_F=4/3$ is
the color factor.  From eq.(\ref{eq:delta_mb}) we immediately see that
the $\Delta m_b$ correction is a finite mass renormalization of the
external bottom legs. The correction as written in
eq.(\ref{eq:delta_mb}) is already resummed over the string of external
one-loop wave function corrections. The authors of
ref.~\cite{delta_mb_resum} have shown that this correction is the
leading term in powers of $\tan \beta$. The reason why this
contribution is usually referred to as non-decoupling is that for
large supersymmetric particle masses in the loop {\sl and for a large
  trilinear mass parameter $A_b$ or higgsino mass parameter $\mu$},
the correction to the Yukawa coupling does not vanish. This is well
understood, since at the one-loop level it couples the `wrong' Higgs
doublet to the bottom quarks. The large one-loop correction does
therefore not mean that perturbation theory breaks down. At the
two-loop level the corrections should be small again. The $\Delta m_b$
factor is not the only non-decoupling contribution in the MSSM either,
as we would expect from the three scalar vertex $\tilde{b}_2
\tilde{t}_1 H^-$, which is again proportional to $m_b (\mu - A_b
\tan\beta)$\cite{charged_nlo_susy}. But the $\Delta m_b$ corrections
for large values of $\tan\beta$ and small values of $A_b$ are expected
to be dominant. This regime is precisely where the charged Higgs boson
search is promising.\bigskip

To estimate how good the leading $\tan\beta$ approximation given by
$\Delta m_b$ is, we also compute the whole set of MSSM loop diagrams.
The result for two different Higgs boson masses is shown in
Fig.~\ref{fig:susy}(a). None of the supersymmetric corrections show a
considerable dependence on the supersymmetric mass scale. To simplify
the presentation we choose a diagonal line in the mSUGRA parameter
space~\cite{drees_martin}: the scalar and gaugino mass scales are
identified $m_{\rm SUGRA}=m_0 \equiv m_{1/2}$\footnote{The
  next-to-leading order calculation is done with a completely general
  MSSM spectrum. The Fortran90 code can be obtained from
  tilman.plehn@@cern.ch.}.  The values for $\tan\beta=30$ and $A_0=0$
are fixed, giving $A_b=0$ at the electroweak scale.  For the $\Delta
m_b$ corrections the sign of the higgsino mass parameter is crucial:
for $\mu<0$ we find $\Delta m_b<0$, which according to
eq.(\ref{eq:delta_mb}) enhances the cross section. For the opposite
sign of $\mu$ the $\Delta m_b$ corrections to the production cross
section are negative. The supersymmetric corrections apart from the
$\Delta m_b$ corrections are negligible in comparison with the $\Delta
m_b$ terms.  This is a feature of the large value of $\tan \beta$ and
is even more pronounced for $\tan\beta=50$ in Fig.~\ref{fig:susy}(b).
We note, however, that the picture changes significantly once we do
not run the higgsino mass parameter $|\mu|$ to large values, together
with the other heavy supersymmetric masses. In that case the $\Delta
m_b$ corrections decouple as shown in Fig.~\ref{fig:susy}(c).
Moreover, for a value $\tan\beta=10$ the $\Delta m_b$ correction drops
below a $\pm 2\%$ effect, becoming even smaller than the explicit MSSM
loop corrections.  We note, however, that choosing large values for
$\tan\beta$ and $|\mu|$ can in principle lead to almost arbitrarily
large $\Delta m_b$ effects, only limited by unitarity constraints.
\smallskip

Heavy particle loops contribute to both the running strong coupling
$\alpha_s(\mu_R)$ and the third generation Yukawa coupling
$y_{b,t}(\mu_R)$. They give rise to supersymmetric counter terms and
thereby yield a logarithmic divergence $\log(m_{\rm heavy}/\mu_R)$ in
the cross section. On the other hand we use Standard Model
measurements for these observables, which means that their running has
to be governed by the light particle beta
function~\cite{lhc_susy_strong,lhc_susy_weak}. The contributions from
heavy particles to their beta function has to be explicitly cancelled,
and as expected this decoupling absorbs all logarithmically
divergences in the one-loop cross section. We show the (misleading)
result one would get without decoupling the heavy particles from the
running Yukawa coupling in Fig.~\ref{fig:susy}(d)\footnote{We note
  that the decoupling in all curves of Fig.~\ref{fig:susy}(a)-(c) is
  computed assuming that all heavy supersymmetric particle masses, the
  gluino mass, the two sbottom masses and the two stop masses, are
  degenerate. This leads to the simple decoupling term $m_{b,t}(\mu_R)
  \to m_{b,t}(\mu_R) [1+\alpha_s/(4 \pi) C_F \log(\mu_R^2/m_{\rm
    heavy}^2)]$. The decoupling term for the strong coupling constant
  is as usually split into contributions from each heavy particle,
  including the top quark~\cite{lhc_susy_strong,lhc_susy_weak}.}.

\section{Summary}

We have computed the complete next-to-leading order contributions to
the inclusive cross section $pp \to tH^-$ in a general two Higgs
double model and in the MSSM. We show why the bottom parton approach
is valid for this process and gives a numerically reliable prediction
for the cross section.

The one-loop contributions hugely improve the theoretical uncertainty
of the leading order cross section prediction, in which one formally
would still have the choice of using a pole mass or a running mass
bottom Yukawa coupling. At next-to-leading order we fix the counter
term to the running Yukawa coupling and check the cross section
dependence on the renormalization and the factorization scale.  Both
lead to an uncertainty of $\lesssim 20\%$ on the total cross
section. The impact on rapidity and transverse momentum distributions
is tested and well under control. The over-all corrections to the
total cross section in the two Higgs double model range between
$+30\%$ and $+40\%$ for Higgs boson masses between $250$ and
$1000\gev$ for the average final state mass scale choice.

In case of a charged Higgs boson in the MSSM two kinds of
supersymmetric corrections appear in addition: the on-shell
renormalization of the bottom quark mass alters the relation between
the bottom mass and the bottom Yukawa coupling. These $\Delta m_b$
corrections are the leading supersymmetric one-loop corrections with
respect to powers of $\tan\beta$. Their effect on the total cross
section in a simple mSUGRA model we estimate to stay below $\pm 5\%$
for $\tan\beta=30$ and below $\pm 20\%$ for $\tan\beta=50$. Because
the charged Higgs boson searches are most promising in the large
$\tan\beta$ regime the remaining explicit supersymmetric loop diagrams
only contribute on a negligible few percent level.

%%%%%%%%%%%%%%%%%%%%  ACKNOWLEDGMENTS  %%%%%%%%%%%%%%%%%%%%

\acknowledgements I would like to thank Dieter Zeppenfeld for numerous
enlightening discussions without which this paper could not have been
written. Michael Spira I would like to thank for carefully reading the
manuscript and being of great help through various interesting
discussions.  Tao Han, Ed Berger, and in particular Jing Jiang I would
like to thank for discussions during an early stage of this project. I
would like to thank Elzbieta Richter-Was and Dave Rainwater for their
curiosity and their encouragement to publish these results.  Dave
Rainwater and Dieter Zeppenfeld I also want to thank for carefully
reading the manuscript.  Last but not least I would like to thank
Shouhua Zhu for the efficient and pleasant comparison of his
results~\cite{charged_zhu} with the ones presented in this paper.
 
This research was supported in part by the University of Wisconsin
Research Committee with funds granted by the Wisconsin Alumni Research
Foundation and in part by the U.~S.~Department of Energy under
Contract No.~DE-FG02-95ER40896. 

%%%%%%%%%%%%%%%%%%%%%%%  REFERENCES  %%%%%%%%%%%%%%%%%%%%%%%

\bibliographystyle{plain}

\end{document}